\documentclass [12pt]{article}
\usepackage{graphics}
\begin{document}                                                                
\baselineskip 18pt
                                                                           
\begin{center}
{\bf
An analytic treatment of the Gibbs-Pareto behavior in 
wealth distribution}  \\
\vspace {0.3 cm}
{Arnab Das and  Sudhakar Yarlagadda } \\
{\it Theoretical Condensed Matter Physics Division and  
Center for Applied Mathematics and Computational Science, 
Saha Institute of Nuclear Physics, Calcutta, India}             
\end{center}
\vspace{0.5cm}
\begin{abstract}
{We develop a general framework, based on Boltzmann transport theory,
 to analyze the distribution of wealth in societies. 
Within this framework we derive the distribution function of wealth
by using a two-party trading model for the poor
people while for the rich people  a new model
is proposed where interaction with wealthy
entities (huge reservoir) is relevant. At equilibrium,
the interaction
with wealthy entities gives a power-law (Pareto-like)
behavior in the wealth distribution while the two-party interaction 
gives a Boltzmann-Gibbs distribution.}
\end{abstract}

\vspace{0.5cm}

\noindent{{\bf PACS classification codes:} 89.65.Gh, 87.23.Ge, 02.50.-r}

\vspace{0.5cm}

\noindent{{\bf Author Keywords:} Econophysics; Wealth distribution; 
 Pareto law; Boltzmann-Gibbs distribution}

%\nopagebreak
%\pagebreak

\section{INTRODUCTION}
 Inequality in the distribution of wealth in the population of a nation        
has provoked a lot of studies.
It is important for both economists
and physicists to
understand the root cause on this inequality: whether
stochasticity or a loaded dice
is the main culprit for such a lop-sided distribution.
While it has been empirically observed by Pareto \cite{pareto}
that the higher wealth group distribution has a power-law tail with exponent
varying  between $2$ and $3$,
the  lower wealth group distribution is exponential or 
Boltzmann-Gibb's like \cite{drag1,mimkes}.
 The Boltzmann-Gibb's law has been shown to be 
obtainable when trading between two people, in the absence of any savings, 
is totally random
\cite{chatchak,adys,drag2}.
 The constant
finite savings case has been
studied earlier numerically by Chakraborti and Chakrabarti \cite{chatchak} and
later analytically by us \cite{adys}. 
As regards the fat tail in the wealth distribution, several researchers
have obtained Pareto-like behavior using approaches such as 
random savings \cite{chat,condmat}, 
inelastic scattering \cite{slanina},  
generalized Lotka Volterra dynamics \cite{solomon}, analogy with
directed polymers in random media \cite{bouchaud}, and three parameter
based trade-investment
model \cite{west}.

 In this paper,
we try to identify the processes that lead to the wealth distribution in
societies. Our model involves two types of trading processes -- tiny and gross.
The tiny process involves trading between
two individuals while the gross one involves trading between an 
individual and the gross-system. The philosophy is that small wealth
distribution is governed by two-party trading while the large wealth
distribution involves big players interacting with the gross-system.
The poor are mainly involved in trading with other
poor individuals.
Whereas the big players mainly interact with 
 large entities/organizations 
such as government(s), markets of nations, etc.
 These 
large entities/organizations are treated as making up the gross-system
in our model. 
The gross-system is thus a huge reservoir of wealth.
Hence, our model invokes the tiny
channel at small wealths while at large wealths the gross
channel gets turned on.

\section{GENERAL FRAMEWORK}
We will now develop a formalism similar to Boltzmann transport theory
so as to obtain the distribution function $f(y,\dot{y},t)$ for wealth $y$, 
net income $\dot{y}$ (or total income after
 consumption) as a function of time $t$.
Similar to Boltzmann's postulate
we also postulate a dynamic law of the form
\begin{equation}
\frac{\partial f}{\partial t}
=\left \{ 
\frac{\partial f}{\partial t}
\right \}_{\rm{ext~source}} +
\left \{ 
\frac{\partial f}{\partial t} 
\right \}_{\rm{interaction}} .
\label{boltz}
\end{equation}
The first term on the right hand side (RHS)
 describes the evolution due to external income
sources only, while the second one represents contribution
from entirely internal interactions.

\subsection{Model for tiny-trading}
  Individuals, possessing wealth
smaller than a cutoff wealth
 $y_c$, engage in two-party trading where two individuals
$1$ and $2$ put forth a fraction of their wealth  
$(1-\lambda _t) y_1$ and $(1-\lambda_t) y_2$ respectively
[with $0 \leq \lambda_t < 1$]. Then the total
money  $(1-\lambda_t)( y_1 + y_2) $ is randomly distributed between the
two. The total money between the two is conserved in the two-party
trading process. We assume that probability of trading by
individuals having certain money is proportional
to the number of individuals with that money.

Since in a two-party trading 
there is no external income
source, then
$\left \{ 
\frac{\partial f}{\partial t} 
\right \}_{\rm{ext~source}} = 0$. The second term
on the RHS in Eq. (\ref{boltz}) can be obtained 
as follows in terms of a balance equation.
\begin{equation}
\frac{\partial f}{\partial t}
=
\left \{ 
\frac{\partial f}{\partial t}
\right \}_{\rm{interaction}} = \rm{gains} - \rm{losses} .
\label{balance}
\end{equation}
In Eq. (\ref{balance}) the two terms on the RHS
can be expressed in terms of transition rates 
$r(y_1 , y_2; y_1^{\prime}, y_2^{\prime})$ 
for a pair of persons to go from moneys $y_1, y_2$
to moneys $y_1^{\prime}, y_2^{\prime}$ respectively.
Then, we have on assuming that the distribution function is only
a function of wealth and time, 
\begin{eqnarray}
\frac{\partial f(y_1,t)}{\partial t}
 && = 
\int
r(y_1^{\prime}, y_2^{\prime};y_1, y_2)
f(y_1^{\prime},t)
f(y_2^{\prime},t)
 dy_2 dy_1^{\prime} dy_2^{\prime}
\nonumber \\
 && - 
\int
r(y_1 , y_2; y_1^{\prime}, y_2^{\prime})
f(y_1,t)
f(y_2,t)
 dy_2 dy_1^{\prime} dy_2^{\prime} .
\label{inter}
\end{eqnarray}
In the above equation, conservation law requires that $y_1+y_2=
y_1^{\prime}+y_2^{\prime}$. Hence in the first integral we treat $y_2$
as redundant and integrate out with respect to it to yield a
normalization constant. Similarly in the second integral
$y_2^{\prime}$ is integrated out.
Now for the transition rate in the first integral of the above equation
we have
\begin{eqnarray}
 r(y_1^{\prime}, y_2^{\prime};y_1, y_2) \propto
 \frac{1}{(1-\lambda_t)(y_1^{\prime} +y_2^{\prime})} ,
\end{eqnarray}
 if
 $\lambda_t y_1^{\prime} \leq y_1 \leq 
 y_1^{\prime}  
+(1-\lambda_t) y_2^{\prime}$  and zero otherwise.
On taking into account the restriction that no one can have negative money
and setting $y_1^{\prime}+ y_2^{\prime} = L$,
the first integral in Eq. (\ref{inter})  at equilibrium is proportional to
\begin{equation} 
\int_{y_1}^{\infty} dL
\int_{a(y_1, L, \lambda_t )}^{b(y_1, L, \lambda_t )}
dy_1^{\prime} 
{\cal{F}}(y_1^{\prime},L, \lambda_t ) ,
\label{Iint}
\end{equation} 
where 
\begin{eqnarray*}
a(y_1, L, \lambda_t ) \equiv \max [0, \{y_1-(1-\lambda_t)L \} /\lambda_t] ,
\end{eqnarray*}
\begin{eqnarray*}
b(y_1, L, \lambda_t ) \equiv \min [L,y_1/\lambda_t ] ,
\end{eqnarray*}
and
\begin{eqnarray*}
{\cal{F}}(y_1^{\prime},L, \lambda_t ) \equiv
\frac{f(y_1^{\prime}) f(L- y_1^{\prime})}{ (1 - \lambda_t ) L} . 
\end{eqnarray*}
As regards the second integral in Eq. (\ref{inter}),
 at equilibrium we assume
that the transition 
from $y_1$ to all other levels is proportional to $f(y_1)$.
Also, since at equilibrium
$\frac{\partial f(y_1,t)}{\partial t} = 0$, we 
obtain the distribution function to be 
\begin{equation} 
f(y)=\int_{y}^{\infty} dL 
\int_{a(y, L, \lambda_t )}^{b(y, L, \lambda_t )}
dx
{\cal{F}}(x,L, \lambda_t ) .
\label{distlamb}
\end{equation} 
The above result was obtained earlier by us using an alternate
 route \cite{adys}.

On introducing an upper cutoff $y_c$ for the two-party
trading, the contribution 
to the  distribution function $f(y)$
from the tiny channel becomes
\begin{equation} 
\gamma \int_{y}^{\infty} dL 
\int_{a(y, L, \lambda_t )}^{b(y, L, \lambda_t )}
dx
{\cal{F}}(x,L, \lambda_t ) 
{\cal{H}}(x,L, y_c ) .
\label{distlambyc}
\end{equation} 
In the above equation
\begin{eqnarray*}
{\cal{H}}(x,L, y_c ) \equiv
[1- \theta(x-y_c )]
[1- \theta(L-x-y_c )] ,
\end{eqnarray*}
with $\theta (x) $ being the unit step function
and $\gamma = 1/\int_{0}^{y_c} dx f(x)$ is a normalization constant
introduced to account for the less than unity value of the
probability of picking a person below $y_c$.

\subsection{Model for gross-trading}
Next, we will analyze the contribution to the
distribution function $f(y)$ from gross-trading.
 An individual
possessing wealth larger than a cutoff wealth ($y_c$) trades with
a fraction $( 1 - \lambda_g ) $ 
 of his
wealth $y$ with the gross-system. The latter
 puts forth an equal amount of money $(1-\lambda_g) y$. 
The trading involves the total sum $2(1-\lambda_g)y$
being randomly distributed between the individual and the reservoir.
 Thus on an average the gross-system's 
wealth is conserved. 

When only the gross-trading channel is operative, we have
\begin{eqnarray}
\frac{\partial f(y_1,t)}{\partial t}
&& = 
\int
r(y_1^{\prime};y_1 )
f(y_1^{\prime},t)
  dy_1^{\prime} 
\nonumber \\
 && 
- \int
r(y_1 ; y_1^{\prime})
f(y_1,t)
  dy_1^{\prime}  ,
\label{resinter}
\end{eqnarray}
where $ r(y_1^{\prime};y_1 )$ is the transition rate from $y_1^{\prime}$
to $y_1$ through interaction with the reservoir.
 Total money involved in trading (between individual and the
gross-system) is $2 y_1^{\prime} (1 - \lambda_g )$. After interaction, the
resulting money $y_1$ of the individual satisfies the
following constraints
 $\lambda_g y_1^{\prime} \leq y_1 \leq (2-\lambda_g) y_1^{\prime}$. 
The first transition rate 
 in the above equation is given by
\begin{eqnarray}
r(y_1^{\prime}, ;y_1) \propto
 \frac{1}{ 2 y_1^{\prime}
(1-\lambda_g) } ,
\label{resrate}
\end{eqnarray}
if $\lambda_g y_1^{\prime} \leq y_1 \leq (2-\lambda_g) y_1^{\prime}$
and zero otherwise.
As before, at equilibrium 
the second integral in Eq. (\ref{resinter})
 is proportional to
$f(y_1)$ when only the gross channel is operative.
 Then the distribution function
$f(y)$ is given by
\begin{equation} 
f(y)=\int_{y/(2-\lambda_g)}^{y/\lambda_g} \frac{dx f(x)}{2 x (1 - \lambda_g ) } .
\label{eqmacy}
\end{equation} 
Now it is interesting to note that the solution of the above equation
is given by $ f(y)=c/y^n $.
To obtain $n$ one then solves the equation
\begin{equation} 
(2-\lambda_g)^{n} - \lambda_g ^ n = 2 n (1 - \lambda_g ) ,
\label{eqn}
\end{equation} 
and obtains $n=1,2$.
Only $n=2$ is a realistic solution because it
 gives a finite cumulative probability.
It is of interest to note that the solution 
 is  {\it independent of $\lambda_g$}.
Also, clearly the distribution function makes sense only for $y > 0$.
 On taking into account an upper cutoff $y_c$,
 the contribution to the distribution function
$f(y)$ from the gross channel is
\begin{equation} 
\int_{y/(2-\lambda_g)}^{y/\lambda_g} \frac{dx f(x)}{2 x (1 - \lambda_g ) } 
\theta (x-y_c) .
\label{macy}
\end{equation} 

\subsection{Hybrid model}

Here an individual
possessing wealth larger than a cutoff wealth $y_c$ does trading with
the gross-system,
while individuals possessing wealth
smaller than $y_c$ engage in two-party tiny-trading.
Hence from Eqs. (\ref{distlambyc}) and (\ref{macy}),
the distribution function
 is obtained to be
\begin{eqnarray}
f(y) = 
&&
\gamma \int_{y}^{\infty} dL 
\int_{a(y, L, \lambda_t )}^{b(y, L, \lambda_t )}
dx
{\cal{F}}(x,L, \lambda_t ) 
{\cal{H}}(x,L, y_c ) 
\nonumber \\
&&
+
\int_{y/(2-\lambda_g)}^{y/\lambda_g} \frac{dx f(x)}{2 x (1 - \lambda_g ) }
\theta (x-y_c) .
\label{distfy}
 \end{eqnarray}

Now, it must be pointed out that when
 the savings 
$\lambda_t =0 $, $\lambda_g \neq 0$, and
%are zero ($\lambda_t =0 $) and
$y \rightarrow 0$,
 Eq. (\ref{distfy}) yields (up to a proportionality
constant) the following same result
as the purely tiny-trading case  without
an upper cutoff \cite{adys}:
 \begin{equation} 
f^{ \prime}(y)
\propto -f(y) f(0) .
\label{distapp}
 \end{equation} 
In obtaining the above equation
we again assumed that the function $f(y)$ and its first
and second derivatives are 
well behaved.
Then the solution for small $y$
is given by
 \begin{equation} 
f(y)
\propto f(0) exp {[-y f(0)]} .
\label{distappsol}
 \end{equation} 
\section{RESULTS AND DISCUSSION}
The distribution function $f(y)$ can be obtained by
solving the nonlinear integral Eq. (\ref{distfy}).
To this end, we simplify Eq. (\ref{distfy}) for computational
purposes as follows:
\begin{eqnarray} 
f(y) = 
&&
 \gamma 
{\cal{G}}(y, \lambda_t , y_c)
\int_{y}^{2y_c} dL 
\int_{a(y, L, \lambda_t )}^{b(y, L, \lambda_t )}
dx
{\cal{F}}(x,L, \lambda_t ) 
{\cal{H}}(x,L, y_c ) 
\nonumber \\
&&
+
[1-\theta ( y- y_{as}) ]
\int_{y/(2-\lambda_g)}^{y/\lambda_g} \frac{dx f(x)}{2 x (1 - \lambda_g ) }
\theta (x-y_c) 
\nonumber \\
&&
+
\theta ( y- y_{as}) f(y_{as}) \frac{y_{as}^2}{y^2} ,
\label{distfy2}
 \end{eqnarray} 
where $
{\cal{G}}(y, \lambda_t , y_c)
\equiv
1-\theta [ y-(2-\lambda_t ) y_c ] $ and $y > y_{as}$ gives the asymptotic
behavior  $f(y) \propto 1/y^2$. In our calculations, we have taken $y_{as}$
 to be at least $20 y_c$ and obtained $f(y)$ for all $y$ less than
 $ 2000$ times the average wealth per person $y_{av}$.
 We solved Eq. (\ref{distfy2}) iteratively by choosing a trial function,
substituting it on the RHS
 and obtaining a new trial function and 
successively substituting the new trial functions over and over again
on the RHS until convergence is achieved.
The criterion for convergence was that
the difference between the
new trial function $f_n$ and the previous trial function  $f_p$
satisfies the accuracy test 
$\sum_i |f_n (y_i) -f_p(y_i)|/\sum_i f_p(y_i) \leq 0.002$ \cite{note}.  

\begin{figure}
\centering{
%\resizebox*{4.25in}{3.125in}{\rotatebox{0}{\includegraphics{fig1.eps}}}}
\resizebox*{4.76in}{3.5in}{\rotatebox{0}{\includegraphics{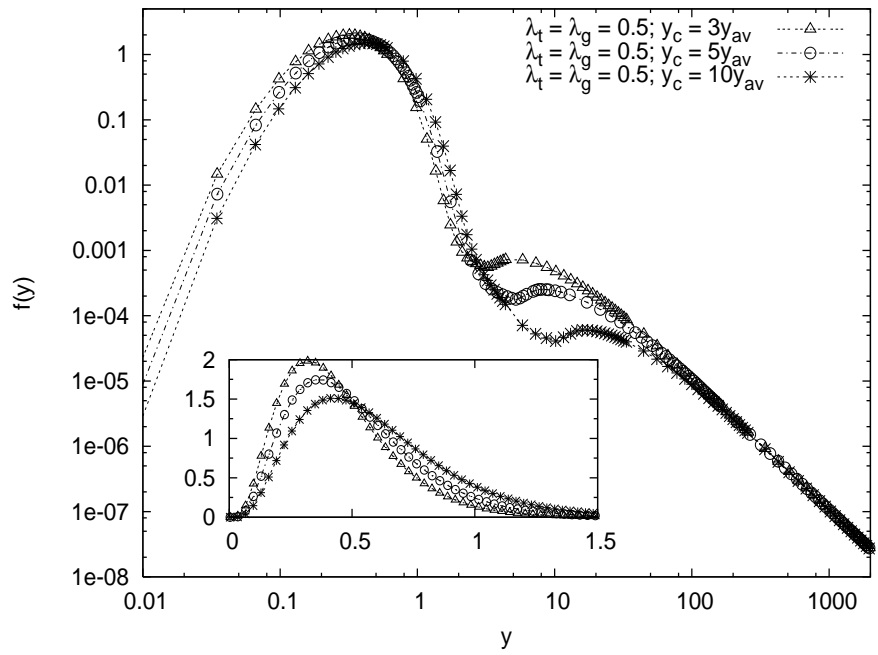}}}}
%\resizebox*{6.8in}{5.0in}{\rotatebox{0}{\includegraphics{fig1.eps}}}}
%\resizebox*{3.4in}{2.5in}{\rotatebox{0}{\includegraphics{fig1.eps}}}}
\vspace*{0.5cm}
\caption[]
{ Plot of the wealth distribution function for savings 
$\lambda_t = \lambda_g = 0.5$
and  various wealth cutoff values $y_c = 3, 5, 10$.
The average money per person $y_{av}$ is set to unity. The dotted lines
are guides to the eye.}
\label{scaling1}
\end{figure}
\nopagebreak
\begin{figure}
\centering{
%\resizebox*{4.25in}{3.125in}{\rotatebox{0}{\includegraphics{fig2.eps}}}}
\resizebox*{4.76in}{3.5in}{\rotatebox{0}{\includegraphics{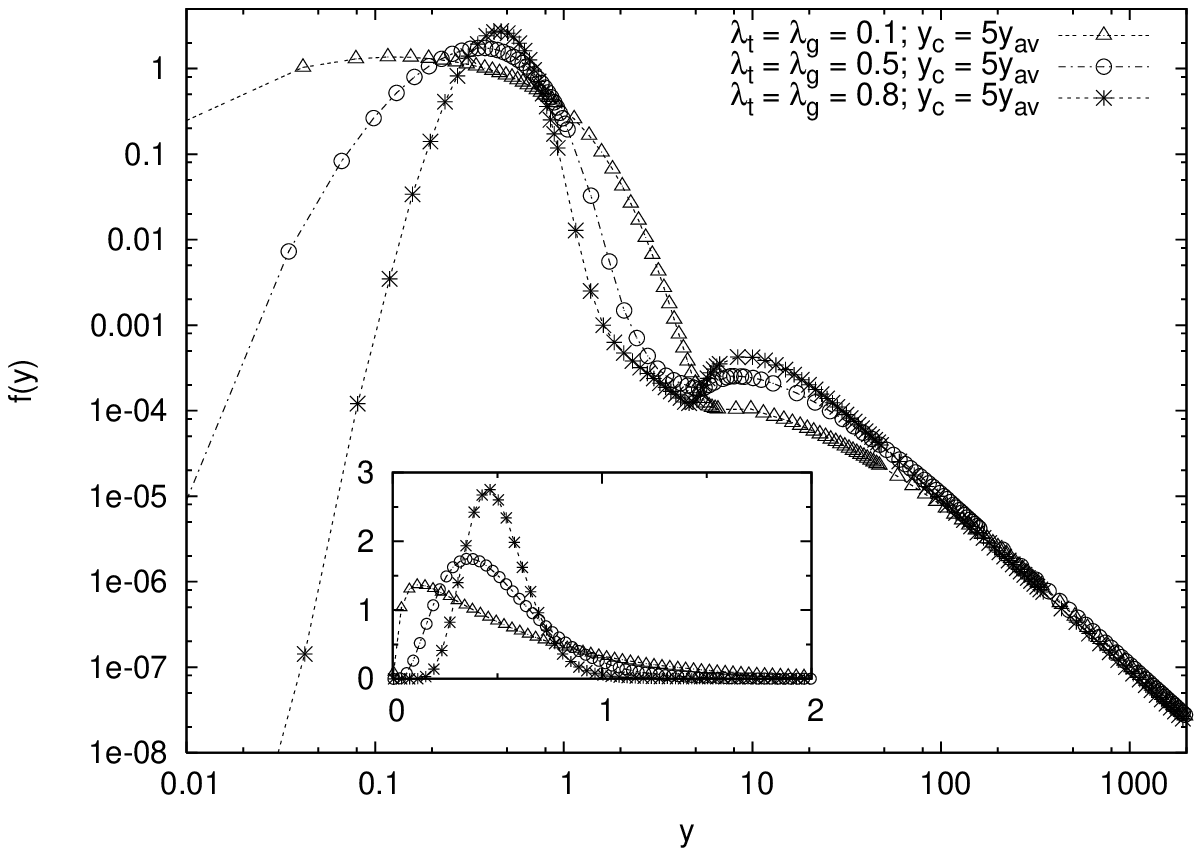}}}}
%\resizebox*{6.8in}{5.0in}{\rotatebox{0}{\includegraphics{fig2.eps}}}}
%\resizebox*{3.4in}{2.5in}{\rotatebox{0}{\includegraphics{fig2.eps}}}}
\vspace*{0.5cm}
\caption[]
{Wealth distribution $f(y)$ at average wealth
$y_{av} = 1$, 
wealth cutoff $y_c = 5$,
and various values of savings $\lambda_t = \lambda_g = 0.1, 0.5, 0.8$.}
\label{scaling2}
\end{figure}
\nopagebreak
In Fig. 1, using a log-log plot we depict the distribution function $f(y)$
for the constant savings case $\lambda_t = \lambda_g = 0.5 $
with the average money per person $y_{av}$ being set to unity and with
the values of the wealth cutoff $y_c = 3, 5, 10$.
As expected, for larger values of $y_c$, the Pareto-like $1/y^2$ behavior
sets in later. The transition to purely gross-trading occurs at
$(2-\lambda_t) y_c$, while below $\lambda_g y_c$ it is purely two-party
tiny-trading. Thus the transition from purely
tiny-trading  to purely gross-trading occurs
 in Fig. 1 over a region of width $y_c$. 
However, all the tails merge irrespective of the cutoff.
At smaller values of $y$ the behavior of $f(y)$, depicted in the inset,
 is similar to the purely
two-party trading model studied earlier (see Ref. \cite{adys}).
The curves in the inset appear to be close because here the 
trading is two-party and is governed by the same savings.
Next, in Fig. 2 we plot $f(y)$ with the cutoff $y_c=5$, $y_{av}=1$, and
for values of savings fraction 
$\lambda_t = \lambda_g = \lambda = 0.1, 0.5, 0.8$.
Here the power-law behavior
($1/y^2$) takes over for $y > (2-\lambda) y_c$ and hence at lower savings
it sets in later. In the power-law region the curves
% more-or-less
merge together. As shown in the inset of Fig. 2,
 at smaller values of $y$ the  $f(y)$s  become zero with
 the higher peaked curves (corresponding to larger $\lambda$s)  
approaching zero faster similar to the case of the purely
two-party trading model in our earlier work \cite{adys}.
Here the transition from purely tiny- to purely gross-trading
at  higher $\lambda$ is sharper because
the transition occurs over a region
of width $2(1-\lambda)y_c$.
\begin{figure}
\centering{
%\resizebox*{4.25in}{3.125in}{\rotatebox{0}{\includegraphics{fig3.eps}}}}
\resizebox*{4.76in}{3.5in}{\rotatebox{0}{\includegraphics{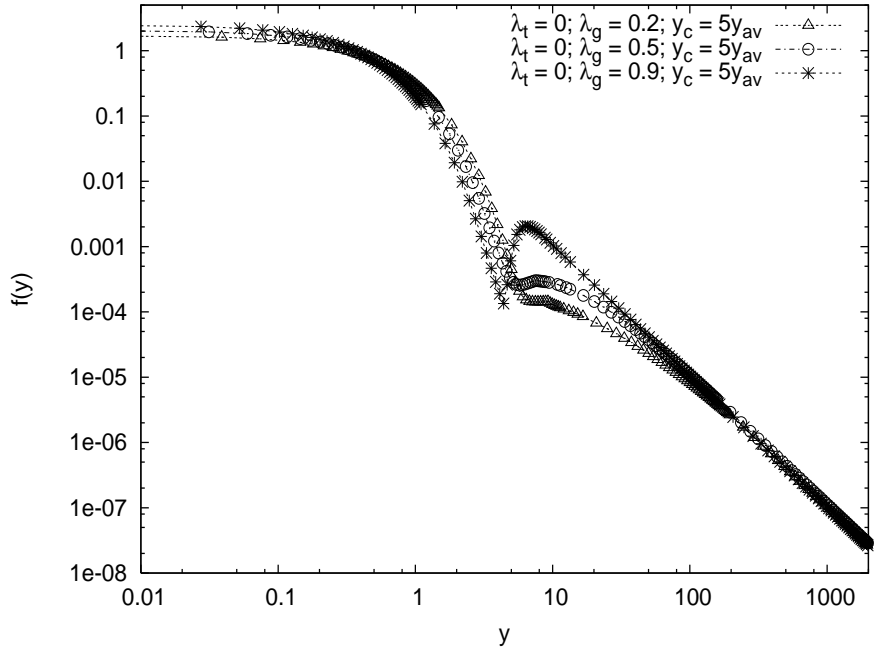}}}}
%\resizebox*{6.8in}{5.0in}{\rotatebox{0}{\includegraphics{fig3.eps}}}}
%\resizebox*{3.4in}{2.5in}{\rotatebox{0}{\includegraphics{fig3.eps}}}}
\vspace*{0.5cm}
\caption[]
{Money distribution function at zero savings for tiny-trading
and various savings values
$\lambda_g = 0.2, 0.5, 0.9$ for the gross-trading.
 The average money 
$y_{av} = 1$ and the wealth cutoff $y_c = 5$. 
}
\label{scaling3}
\end{figure}
\nopagebreak
Lastly, in Fig. 3, we show the distribution function $f(y)$
for the zero savings case in the tiny-channel ($\lambda_t = 0 $)
and for various savings $\lambda_g = 0.2, 0.5, 0.9$ in the
gross-channel
with 
$y_{av} = 1$ and $y_c =  5$.
 The distribution, as expected, decays exponentially 
(or Boltzmann-Gibbs-like) for small values
of $y$ and has power-law ($1/y^2$) behavior at large values.
The curves merge in the Pareto-like region and, in fact, 
 $f(y) \approx 0.1/y^2$
in all  the three figures at large values of $y$.
%with the curves merging in the Pareto-like region.
In Fig. 3 too, for reasons mentioned earlier, the transition is sharper at 
larger values of $\lambda_g$.
Fig. 3 takes into account the fact that, in societies, the rich
tend to have higher savings fraction ($\lambda$)
 compared to the poor.
 Actually, if the savings fraction were to
increase gradually with wealth, one can expect a more gradual
change in the
transition region of
the distribution rather than the sharp local maxima (around $y \approx
6.5$ ) shown by the $\lambda_g = 0.9$ curve.

In all the figures anomalous looking kinks/shoulders appear at the cross over
between the Boltzmann-Gibbs-like and the Pareto-like regimes. This is due to
the sharp cut-off at $y_c$ that we introduced using a step function.
However, a kink seems to be generic in these kind of
distributions in real populations (as borne out by the empirical data
in Fig. 9 of Ref. \cite{drag1})
indicating that two different dynamics may be operative in 
the two regimes.
 Different societies have the
onset of Pareto-like behavior at different wealths which is
indicative that the cut-off has to be obtained empirically 
based on various factors like the social structure, welfare 
policies, type of markets, form of government, etc.

      In Japan the wealth/income distribution vanishes
 at zero wealth/income and then rises to a maximum (see Ref. \cite{drag1}).
In US the distribution
 seems to be a maximum at zero wealth/income (see Ref. \cite{drag1}).
 Both these aspects can be covered in our model as the poor in
 general are known to save very little. If their savings
are zero, one gets the 
  Boltzmann-Gibbs behavior
at the poor end. On the other hand,
 if the savings are
small one gets a maximum close to zero and the distribution vanishes
at zero wealth. 

It would be interesting to deduce
the savings pattern
from the wealth distribution.
While it has been observed that the rich tend to save more than
the poor, 
 how gradually the savings change as wealth increases
can perhaps be inferred from the change in slope.  However, as
explained below, the middle region (involving the middle-class)
 has been modeled quite crudely by us and needs to be refined
before a serious connection with savings pattern can be attempted. 

We will now further discuss the motivation for using two different
mechanisms to model the observed wealth distribution.
     The model is an approximation 
where the direct wealth exchange occurs between people who are in
%of people whose economic interaction is with people within their 
economic proximity.
At the poorer end of the spectrum, the poor,
who have limited economic means and avenues,
 come in contact with a few  poor and their
economic activity is modeled in terms of two-party trading.
At the other end of the wealth spectrum, the rich
 have  access to various economic avenues
(e.g., markets, know-how,
work force, capital, credit facilities, contacts,
wealthy society, etc.) due to which they
 can trade with huge organizations and are thus modeled to interact with
a reservoir.
As regards the middle-class that is between the rich and the poor,
they trade amongst themselves as well as with the  poor and the
reservoir. 
As a first step towards realizing 
this scenario, we included in our model
only the two extreme cases of interaction. What we have not 
taken into account is the interaction of the middle class with 
 the reservoir. To rectify this,
in future we hope to introduce
a cutoff $y_g$  for the interaction with the reservoir
such that $y_g$ lies below the two-party 
trading cutoff $y_t$.
  Thus, we believe that our model is a reasonable  
one at the poor and rich ends and is a crude
 approximation for the middle class. 
 While it is true
that the poor also come in market contact with wealthy organizations
like the coke company, the contact is an indirect one through intermediaries.
For example, the poor person deals with a richer shop-keeper selling coke
who in turn deals with a richer local distributor who in turn deals
with the big coke company. 
Lastly, we would like to add that the assumption
 of random distribution in two-party trading
is a model studied by others as well (see Refs. \cite{chatchak,chat}).
 We feel that in any 
trading there is random fluctuation of the price around its true value.
The total money put forth for trading corresponds to the amount
of random fluctuation. However the poorer of the two  puts forth less
and makes the trading biased in his/her favor. This can be justified
from the fact that the poor people
 are constantly looking for bargains to
 make ends meet.

 Compared to other types of analyses involving two-party trading
to explain Pareto law (see Refs. \cite{chat,condmat}),
 our gross-trading mechanism
can make contact with
the standard approach in macroeconomics  as will be shown below.
Over the past, economists have developed two models, namely, the
dynastic model and the life-cycle model, to explain wealth
distribution. In the dynastic model,
where bequests are vehicles of transmission of wealth
inequality, people save to improve the consumption of their descendants.
On the other hand, in the life-cycle model, where wealth of an individual
is a function of the age,
people save to  provide for their own consumption after retirement. 
Both these models and their hybrid versions have had only limited
success \cite{quadrini}. However, one of the ingredients that
goes into these models, i.e., uninsurable shocks or stochasticity
in income,  has been
exploited by econophysicists with remarkable success
in reproducing power-law tails. 

In macroeconomics, the objective
is to maximize a cumulative utility function 
 subject to a wealth constraint \cite{favero}. Mathematically
this is formulated as
\begin{equation}
\max_{c_{t+i}, y_{t+i}} E_t \sum_{i} \beta^i u(c_{t+i}) ,
\end{equation}
subject to the constraint
\begin{equation}
 y_{t+i} = (1+r) y_{t+i-1} + e_{t+i} -  c_{t+i} ,
\end{equation}
where $c_t$, $y_t$,  and $e_{t}$ are consumption, wealth, and labor earnings
 respectively at time $t$, $r$ is the interest rate on wealth $y$,
$0 < \beta < 1$ is the time-discount factor, $u(c_t)$ is the concave
utility function, $E_t$ is the expectation value based on the available
information at time $t$. Using the method of Lagrange multipliers,
the conditions of optimality yield
\begin{equation}
E_t [ u^{\prime}(c_{t}) - (1+r) \beta u^{\prime}(c_{t+1})]= 0 ,
\label{optut}
\end{equation}
 where $u^{\prime}(c_t)$ is the derivative of $u(c_t)$ with respect
to $c_t$. From the above equation we see that consumption at different
times are related.
In our work [see Eq. (\ref{resrate})], we introduced the stochasticity
\begin{equation}
 y_{t+1} - y_{t} = \epsilon (1- \lambda_g ) y_{t}  ,
\label{stochy}
\end{equation}
 where $\epsilon$ is a random number with
$-1 \le \epsilon \le 1$, which implies that 
\begin{equation}
r y_{t} + e_{t+1} -  c_{t+1} = \epsilon (1- \lambda_g ) y_{t}  .
\end{equation}
The above equation can be made consistent with the optimal consumption
relation given by Eq. (\ref{optut}). Thus our results can
be approached through the standard machinery in macroeconomics.

The stochasticity
in wealth given by Eq. (\ref{stochy}) implies that the spread
in wealth distribution at time $t+1$ is proportional to $y_t$ and thus
 wealth's $y_{t+1} > y_t$
yield a wider spread for $y_{t+2}$  than 
 do wealth's $y_{t+1}  <  y_t$. Thus the distribution becomes
more skewed to the right.

In conclusion, we introduced a new ingredient -- interaction of the
rich with huge entities -- and obtained a Pareto-like
power-law.
On the other hand, the Boltzmann-Gibbs-like wealth distribution
of the  poorer part of the
society is explained through a two-party trading mechanism.
All in all, we demonstrate that stochasticity can account for the observed
skewness in the wealth distribution.

\begin{center}
{\bf ACKNOWLEDGMENTS}
\end{center}

 The authors are grateful to B. K. Chakrabarti for 
 useful discussions.


\begin{thebibliography}{999}
 \bibitem{pareto}
V. Pareto, {\it Cours d'Economie Politique},
(Lausanne, 1897).

 \bibitem{drag1}
  A. A. Dragulescu, cond-mat/0307341.
% A. A. Dragulescu and V. M. Yakovenko, cond-mat/0103544;

 \bibitem{mimkes}
  G. Willis and J. Mimkes, cond-mat/0406694.

 \bibitem{chatchak}
  A. Chakraborti and B. K. Chakrabarti, Eur. Phys. J. B {\bf 17},
 167 (2000).                           

\bibitem{adys}
 A. Das and S. Yarlagadda,
 Physica Scripta T {\bf 106}, 39 (2003).

 \bibitem{drag2}
  A. A. Dragulescu and V. M. Yakovenko, Eur. Phys. J. B {\bf 17},
 723 (2000).                           

 \bibitem{chat}
  A. Chatterjee, B. K. Chakrabarti, and S. S. Manna,
 Physica Scripta T {\bf 106}, 36 (2003).

 \bibitem{condmat}
  A. Chatterjee, B. K. Chakrabarti, and S. S. Manna, 
Physica A {\bf 335}, 155 (2004).

 \bibitem{slanina}
  F. Slanina, 
 Phs. Rev. E {\bf 69}, 046102 (2004).

 \bibitem{solomon}
  S. Solomon and P. Richmond, 
 Physica A {\bf 299}, 188 (2001).

 \bibitem{bouchaud}
J.-P. Bouchaud and M. M\'{e}zard, Physica A {\bf 282}, 536 (2000).

 \bibitem{west}
  N. Scafetta, S. Picozzi, and B. J. West, cond-mat/0403045.

 \bibitem{note}
In Fig. 2, only for the $\lambda_t =\lambda_g = 0.8$ curve,
we choose
$\sum_i |f_n (y_i) -f_p(y_i)|/\sum_i f_p(y_i) \leq 0.008$
to cut down computational time.  

 \bibitem{quadrini}
V. Quadrini and J.-V. R\'{\i}os-Rull, Federal Reserve Bank of Minneapolis
 Quarterly Review, {\bf 21 (2)}, 22 (Spring 1997).

 \bibitem{favero} C. A. Favero, {\em Apllied Macroeconometrics}
(Oxford university press, New York, 2001).

\end{thebibliography}
\end{document}